# Astrophysical Ionizing Radiation and the Earth: A Brief Review and Census of Intermittent Intense Sources

Adrian L. Melott[1] and Brian C. Thomas[2]

1. Department of Physics and Astronomy, University of Kansas, Lawrence, Kansas 66045 USA. melott@ku.edu, Phone: 785-864-3037. Fax 785-864-5262
2. Department of Physics and Astronomy, Washburn University, Topeka, Kansas 66621 USA. brian.thomas@washburn.edu

**Keywords:** radiation, gamma-ray burst, supernova, solar flare, ozone depletion, extinction, cosmic ray, X-ray.

### Abstract

Cosmic radiation backgrounds are a constraint on life, and their distribution will affect the Galactic Habitable Zone. Life on Earth has developed in the context of these backgrounds, and characterizing event rates will elaborate the important influences. This in turn can be a base for comparison with other potential life-bearing planets. In this review we estimate the intensities and rates of occurrence of many kinds of strong radiation bursts by astrophysical entities ranging from gamma-ray bursts at cosmological distances to the Sun itself. Many of these present potential hazards to the biosphere: on timescales long compared with human history, the probability of an event intense enough to disrupt life on the land surface or in the oceans becomes large. Both photons (e.g. X-rays) and high-energy protons and other nuclei (often called "cosmic rays") constitute hazards. For either species, one of the mechanisms which comes into play even at moderate intensities is the ionization of the Earth's atmosphere, which leads through chemical changes (specifically, depletion of stratospheric ozone) to increased ultraviolet-B flux from the Sun reaching the surface. UVB is extremely hazardous to most life due to its strong absorption by the genetic material DNA and subsequent breaking of chemical bonds. This often leads to mutation and/or cell death. It is easily lethal to the microorganisms that lie at the base of the food chain in the ocean. We enumerate the known sources of radiation and characterize their intensities at the Earth and rates or upper limits on these quantities. When possible, we estimate a "lethal interval," our best estimate of how often a major extinction-level event is probable given the current state of knowledge; we base these estimates on computed or expected depletion of stratospheric ozone. In general, moderate level events are dominated by the Sun, but the far more severe infrequent events are probably dominated by gamma-ray bursts and supernovae. We note for the first time that so-called “short-hard” gamma-ray bursts are a substantial threat, comparable in magnitude to supernovae and greater than that of the higher-luminosity long bursts considered in most past work. Given their precursors, short bursts may come with little or no warning.

## I. Introduction

The Earth is continually bombarded by radiation from the rest of the Universe. It comes in many varieties, intensities, and is subject to great variation. This of course includes the gentle starlight and other sources of information about the rest of the Universe, as well as the sunlight that fuels the biosphere. Some of the radiation can be dangerous. This is particularly true if an energetic event is unusually close, energetic, or pointed at the Earth. Most of the time the biosphere is protected by the Earth’s atmosphere and magnetic field, which absorb and channel much of the potentially damaging radiation. Travelling outside the atmosphere, and especially outside the magnetic shield that extends to low-Earth orbit, renders astronauts vulnerable to Solar storms. Nearly all the individual radiation sources that affect us are subject to bursts or possible long-term enhancement that may make them dangerous. Over geological timescales, this becomes an interesting question for the development of the biosphere. Although we have been observing such things directly for only a short time, we can summarize the observations and use indirect arguments to estimate or set limits on the rates of many such events. It is the purpose of this review to summarize what is known at this time, in terms of rates and intensities, as well as point out some areas of further useful research. This review is aimed at a wide audience from a range of fields, and so we include some background material that will be obvious to some readers but new to others.

It is beyond the scope of this document to discuss in more than a superficial way the biological effects of the various types of radiation. The effects of some important kinds are covered in texts (e.g. Alpen 1997; Kudryashov and Lomanov 2008). Others, such as high-energy muons, are important in air showers, but relatively less attention has been paid to understanding their biological effects. One well-known effect is mutation (Moeller et al. 2010). It is important to note that an increase in the mutation rate does not necessarily lead to more rapid evolution; the limiting factor is often something else, such as selection pressure or isolation of small populations of a species. Other effects involve carcinogenesis and the inhibition of photosynthesis. The atmosphere may be changed in a way that admits more of certain types of damaging ultraviolet light, as discussed below.

We need a semiquantitative way to characterize the probable damage level to the biota of various sources over geological intervals. Given that the ozone depletion/UVB effects are potentially severe, probably the most easily triggered, and relatively well-studied, we will use them as calibration. Given that current anthropogenic depletion is doing measurable damage to our biosphere (Häder 1997; Häder et al. 2007), we use the threshold of "measurable damage" as that which produces a mean global ozone depletion of about 3%, close to that recently observed. A mean depletion of about 30% will nearly double the mean UVB flux at the surface (Thomas et al. 2005). This is far above the level of lethality for phytoplankton in particular, and would be expected to trigger a food chain crash in the ocean (Melott and Thomas 2009), so we will call this an "extinction level event." These definitions are not precise, because mortality will vary between different organisms, and its consequences will depend upon what else is going on in the biosphere. They are, however, sufficiently precise given the level of our astrophysical knowledge about source rates and intensities.

## II. Characterizing types of ionizing radiation

We consider here both electromagnetic radiation (photons) and high-energy protons and other nuclei (often called "cosmic rays"). We will label electromagnetic radiation by its photon energy, which is proportional to frequency (and therefore inversely proportional to wavelength). Strong effects on molecules happen when chemical bonds can be broken, which means roughly a few eV and up. The eV is a unit of energy corresponding to that gained by one electron when it falls through a potential difference of 1 volt. Modern astronomy has gone far beyond visible light, and uses all electromagnetic radiation as windows on the Universe—often from orbiting observatories. Photons are immune to the effects of magnetic fields but can be stopped by matter. Most of the energetic photons do not penetrate very far into the Earth's atmosphere, but they can change it.

The other main type of radiation (usually called "cosmic rays") consists of elementary particles accelerated to very high speeds, i.e. close to that of light. The most important cosmic rays are those particles which are the constituents of ordinary matter. The cosmic ray as it arrives at the upper atmosphere is called a primary; it will undergo many interactions and rarely arrives unchanged at the surface of the Earth. The vast majority of primaries are protons, the positively charged nuclei of hydrogen atoms, the most common element in the Universe. Alpha particles, the nuclei of helium atoms with 4 times the mass of protons are an important subcomponent; the rest are electrons and the nuclei of heavier elements. Recall that mass and energy are convertible; it is customary to refer to the "rest mass" of a proton as the energy into which it might be converted, about 1 GeV. Lower-energy primaries, especially those from the Sun, might be keV. This means they are carrying about a keV of kinetic energy, far less than their "rest mass," which implies they are moving at speeds far less than the speed of light. At higher energies, their kinetic energy far exceeds their rest mass, and they are moving very close to the speed of light. High energy cosmic rays are reviewed in depth by Kotera and Olinto (2011).

Sources of cosmic rays include our Sun, both as a steady source and occasional bursts in solar flares. The so-called anomalous cosmic rays come from the outer Solar System, where there is a shock front between the outgoing "solar wind" and the interstellar medium. Supernovae are thought to produce cosmic rays up to $10^{15}$ eV (e.g. Caprioli, Blasi, and Amato 2010). As there are a few supernovae per century in our galaxy, and the travel times are large (see below), the cosmic ray background is fairly constant unless one goes off relatively nearby (Erlykin and Wolfendale 2006, 2010). Highly energetic cosmic rays, up to about $10^{21}$ eV are thought to be produced by active galactic nuclei (Abraham et al. 2007) and/or gamma-ray bursts (Dermer and Holmes 2005; Calvez et al. 2010).

The path of charged particles in a uniform magnetic field is typically a spiral or helix, following around the "field lines." In the case of the Earth, where field lines typically connect the poles, this channels most charged particles to the polar regions, so that residents there are exposed to more radiation. When the energy gets up to about 17 GeV or more, charged nucleons basically punch through the Earth's magnetic field and interact with the atmosphere all over the globe (Usoskin and Kovaltsov 2006). Much of the energy is deposited in the atmosphere. Rarely do primaries reach the ground; instead secondary particles of various kinds dominate what is received there. Additional information on cosmic rays is summarized in Ferrari and Szuszkiewicz (2009).

## III. **Dominant terrestrial effects of ionizing radiation**

In this paper we describe "measureable", "lethal" or "extinction level" events, in terms of their effect on the biosphere. There are several ways that photon and particle radiation from astrophysical sources can affect life on Earth. Here we divide these effects into two main categories: direct and indirect. For many of the sources we consider, the most important impact is an indirect one – ionization of the atmosphere leading to depletion of stratospheric ozone. As described below, this depletion leads to biological damage due to solar UVB radiation. We therefore use ozone depletion as a proxy for biological damage (and hence how "dangerous" an event may be).

It is beyond the scope of this summary to extensively discuss terrestrial effects of episodes of enhanced ionizing radiation. Increased transmission of damaging Solar UVB is expected to be the most important effect triggered by an ionization event. There is extensive literature in photobiology, particularly on the effects of UVB since somewhat increased UVB from anthropogenic changes in the atmosphere has been a concern in recent decades; we cite some of this below.

Gamma-ray bursts (the most energetic events considered here) emit X-ray and gamma-ray photons. These are nearly all absorbed by the atmosphere. The important effects are changes in the chemistry of the atmosphere from ionization by this radiation, and to a lesser extent a "flash" of secondary photons which reach the ground. Such photons are also important to consider with other radiation sources.

Although cosmic rays also ionize the atmosphere (Atri et al. 2010; Melott et al. 2010a), in doing so they typically disappear and set off a shower of secondary particles. Unlike astronauts, who may be exposed to the primaries, the secondary shower is most important for life on the ground. Computation of the contents of air showers from high-energy primaries is only now getting underway (e.g. Atri and Melott 2010).

The only certain diagnostic of past atmospheric events is the formation of unstable isotopes by nuclear interaction of cosmic rays with components of the atmosphere. Most prominent are $^{14}C$, $^{10}Be$, and $^{26}Al$, with half–lives of 5,370 yr, 1.51 Myr, and 717,000 yr respectively. These are steadily produced in the atmosphere but are enhanced during episodes of increased cosmic ray flux. They can be found in biological samples, speleothems (such as stalagmites) and ice cores. With their geologically short decay times, they unfortunately cannot probe the long timescales needed to diagnose the infrequent strong events likely to be associated with supernovae and gamma-ray bursts.

Cosmic ray effects on the biosphere include direct radiation and secondaries (e.g. Karam 2003). There is considerable statistical evidence linking even the greatly attenuated effects of cosmic rays on the ground with increased cancer and birth defects (e.g. Juckett and Rosenberg 1997; Juckett 2007, 2009). There have been attempts to probe the fossil record for bone cancer effects as a proxy for radiation episodes (Natarajan et al. 2007). Bone cancer can be caused by thermal neutrons, which have a flux of only a few per $cm^2$ per second on the ground, but about 280 times higher in the stratosphere, where they are probably responsible for elevated cancer rates seen in airline attendants (Goldhagen 2003). Such an effect could reach the ground with more energetic cosmic rays. It is unlikely with the kind of irradiation expected from many astrophysical sources, but is very likely with the kind of cosmic rays likely to be incident from a nearby supernova (Erlykin and Wolfendale 2010) or possibly from a gamma-ray burst in our galaxy (Dermer and Holmes 2005). We have work in progress to compute the thermal neutron flux on the ground from high-energy cosmic rays from such events. Treatment of cosmic ray effects is complicated by the fact that they are deflected in the magnetic field that permeates the galaxy. Only for the extremely high energy protons that may come from a gamma-ray burst are the trajectories likely to be ballistic; for most cosmic rays the propagation is effectively diffusive. A planet in the path of the beam would receive an initial burst of directional high-energy cosmic rays, followed by a progressively more extended influx of lower-energy cosmic rays which would appear (due to the same deflection which makes the propagation diffusive) to come from all directions. An approximate model, likely to incorporate the most important effects is presented in Erlykin and Wolfendale (2010). They argue that periodic increases in the flux of PeV cosmic rays are likely, and would trigger great increases in the rate of lightning.

Photons in a narrow range of wavelengths (roughly 380 to 750 nm) are called visible light. A continuous source with a high blue light (photons with wavelengths at the short end, and hence high energy, end of the visible range) content or higher frequency photons retransmitted as blue light might have some deleterious effects on animals, as discussed in Thomas et al. (2008). The mechanism seems to be that there are non-visual channels by which irradiation with blue light can actually change hormone balances in animals (e.g. Brainard and Hanifin 2005; Vandewalle et al. 2007). A source that is bright in the sky at night for weeks or months, such as a nearby supernova, might have considerable effect via this largely unexplored channel.

Another “nearly direct” effect results from the fact that high-energy photons incident on the upper atmosphere are re-transmitted and typically reach the surface as UV light of various frequencies (Smith et al. 2004; Martín et al. 2010; Peñate et al. 2010). UVB (wavelengths 320 nm–280 nm, or photon energies of 3.94–4.43 eV) is particularly damaging to DNA and proteins. The UVB transmission for a gamma-ray burst for an atmosphere like that of the Phanerozoic Earth (roughly the last 500 Myr, which has a good fossil record and is the only time considered in this review) is about 0.0006 of the total energy—which for the canonical extinction-level burst of Thomas et al. (2005) would constitute 60 $J/m^2$ of UVB delivered over 10 seconds. This would about double the flux normally incident at noon on a sunny day in equatorial latitudes. Such an event is probably at most a cause for temporary blindness.  As pointed out by Martin et al. (2010), this direct effect is probably much more important for the early Earth, before the advent of the ozone shield. Both this direct UVB effect and the blue light hormonal disruption mechanism mentioned in the previous paragraph would only be of concern for one hemisphere of the Earth.

The indirect effect of ionization of the atmosphere (e.g. Usoskin et al. 2009) probably dominates over the direct ones in importance for stress on the biosphere in the Phanerozoic, since life has evolved in an environment protected from UVB during this time. Atmospheric effects from ionizing radiation may result in a fairly long-term increase in the transmission to the surface of the Earth of UVB from the Sun. Substantial increase in UVB will persist for several years after the ionization source stops. Useful discussions include Reid (1978), Cockell (1999), and Rothschild (2007). A sketch of the mechanism is appropriate. Ionizing radiation breaks the strong chemical bond of $N_2$, which composes about 80% of the atmosphere. Once the bond is broken, all kinds of oxides of nitrogen are formed—the crucial ones being NO and $NO_2$.  These dissociation/ionization products are normally quite rare in the atmosphere, which explains why nitrates are the most common kind of agricultural fertilizer. These compounds (NO and $NO_2$) set up a catalytic cycle in which ozone ($O_3$) is converted to ordinary oxygen ($O_2$). Stratospheric ozone is the main absorber of UVB photons, and normally this prevents most of the solar UVB radiation from reaching Earth's surface. Depletion of stratospheric ozone therefore results in a strong increase in UVB at the ground. The process is described in Thomas et al. (2005) in some detail. There are a number of interesting systematic effects.  An ionizing radiation event in the autumn of a given hemisphere has the strongest long-term effect. The long-term maximum ozone depletion (and rate of recovery) depends not so much on the rate or duration of irradiation, as on its spectrum and total amount incident upon the Earth (Ejzak et al. 2007). Although east-west atmospheric mixing is rapid, mixing across the equator is not (at least under climatic conditions similar to the present), and an event that irradiates only the northern or southern hemisphere results in depletion largely confined to that hemisphere. We have suggested that a radiation event may have contributed to the end-Ordovician mass extinction event (Melott et al. 2004), and have described a prediction for yet unstudied parts of the fossil record based on the hemispheric asymmetry (Melott and Thomas 2009).

Part of the importance of loss of the ozone shield is based on the catastrophic effects of UVB on protein and on the DNA molecule. These are especially severe for unicellular and other small life forms, since they cannot shield themselves, being essentially transparent. Phytoplankton are responsible for about half the photosynthetic activity on the planet, and are at the base of the food chain in the ocean. Since they are dependent upon sunlight, they must also be exposed to some UVB in a crisis event, which could lead to a food chain crash in the oceans (Melott and Thomas 2009). We are working now to characterize the effect on *Synechococcus* and *Prochlorococcus*, two picophytoplankton species that are abundant in the oceans and thereby make up a large fraction of the primary productivity. A wide variety of UV side effects are important (e.g. Rothschild 2007), producing stress on biota even at the modest UVB enhancement level caused by modern anthropogenic ozone depletion (Häder 1997; Häder et al. 2007).

Any kind of ionizing radiation sufficiently energetic to break the $N_2$ bond can initiate such an event, such as a nearby supernova (Ruderman 1974; Gehrels et al. 2003) or large Solar flare (Thomas et al. 2007). Even bolide impact events can produce similar chemical effects (e.g. Reid 1978; Melott et al. 2010b, Pierazzo et al. 2010). Many major past impact events could be discerned by craters. In the case of cometary airbursts in the last Myr or so, the simultaneous presence of nitrate and ammonium in ice cores should be a clear marker (Melott et al. 2010b). Since impacts have been relatively much more extensively studied than ionizing radiation events, we will restrict our attention to the latter.

The early Earth was of course subject to the effects of cosmic ray secondaries. Although there was very little free oxygen and therefore no significant ozone shield, it has been argued that the early Earth was probably protected against UVB by organic haze (e.g. Wolt and Toon 2010), so organisms should have been protected most of their evolutionary history. During the oxygenated era, however, this protection was subject to damage by radiation (as described above). Consequently, we restrict our discussion of such ionization effects to roughly the last 500 Myr, when loss of the UVB shield could result from atmospheric ionization.

The formation of the Earth ~4.5 Gyr ago is a significant fraction of the age of the Universe in the past. Consequently, based purely on cosmic evolution, we would expect supernova and gamma-ray burst rates to be somewhat higher, but this is not a large effect. The normal background of solar cosmic rays would also be higher (Karam 2003). Most notably, the background from the decay of radioisotopes on the ground was about five times higher (Karam and Leslie 1999). Since the backgrounds on the early Earth were very much larger, the perturbations due to only modestly increased event rates would be less important than in the last 500 Myr, which has less background from radioactive decay and also is accustomed to the ozone shield. Consequently we conclude that the kind of processes we consider here were much less important for the early Earth.

## IV. **Characterizing astrophysical sources of ionizing radiation**

There is of course a background of ionizing radiation on the Earth due to radioactive decay (slightly increased in recent decades as a consequence of nuclear testing). There is also exposure to humans due to medical X-rays, etc. We do not consider these approximately steady backgrounds, except as a base from which to consider possible increases. A good discussion of local terrestrial backgrounds can be found in Karam and Leslie (1999).

Cosmic rays and ionizing electromagnetic radiation (photons) bombard the Earth from various extraterrestrial sources. It is the primary purpose of this review to consolidate and review information on such sources, and in particular their possible large fluctuations over geologic time. Such fluctuations have long been considered a probable intermittent trauma to the biosphere, and a possible causal agent for mass extinctions (e.g. Ruderman 1974). Obviously, extinction can refer to the termination of an individual or species, but mass extinctions have been defined as constituting sudden and widespread extinction of a substantial fraction of species (Bambach 2006). Erlykin and Wolfendale (2006, 2010) have recently considered variations in cosmic rays, with an emphasis on higher-energy ones, about 1-10 PeV.

This document will be primarily organized around generating a “census,” or source list, with an attempt to characterize each candidate source in terms of intensity and frequency of irradiation enhancements, to the extent possible based on both direct and indirect measurements. For the more infrequent and intense events, the evidence becomes more indirect, or we may only be able to characterize what is expected or describe upper limits on the rate of events.

*Solar descreening events* can happen when the Solar System encounters a denser region of the interstellar medium as it moves through that medium. The outflowing “solar wind,” a plasma with embedded magnetic field, normally protects the Earth from the effects of galactic cosmic rays. It may be pushed back so that the Earth is exposed to the full cosmic ray flux as well as possible embedded dust, etc. in the interstellar medium. We will look briefly at this possibility. Such descreening is also an operative mechanism in the case of a blast wave from a very nearby supernova (Fields et al. 2008; Athanassiadou and Fields 2011).

The primary extraterrestrial source of radiation is, of course, the *Sun*, irradiating the Earth with about 1380 watts per square meter. Attenuation by the atmosphere reduces this to about 1000 watts per square meter on a horizontal surface when the Sun is directly overhead on a clear day. Obviously clouds, varying water vapor, dust and other particulates, and many other things can further reduce this. Most of this radiation is in the form of visible light. Organisms are tuned to see the dominant radiation emitted by the Sun. Its spectrum is shaped approximately like a blackbody spectrum at 5250 °K. However, the Sun also produces ionizing radiation, in the form of low to moderate energy cosmic rays, ultraviolet light, and X-rays. These can greatly increase during short times, for example during Solar flares. In addition, the Sun was more consistently active in the past (Karam and Leslie 1999; Karam 2003).

*Supernovae* are spectacular explosions that take place at the end of the “normal life” of some stars (depending upon mass). There is a detailed taxonomy, because there are a number of types of supernovae, with different kinds of detailed behavior. Some of them result when the core of a massive (more than at least six Solar masses) star runs out of fuel and collapses, producing an explosion. Others result when a stellar remnant called a white dwarf accretes mass from its companion in a binary system, and exceeds the limit of stability, with an explosion resulting. We will not dwell on this, except in order to characterize the kind of radiation the Earth would be likely to receive. The rate of supernovae is something like 3 per century in our Galaxy; most of these will be invisible due to obscuration. Our Galaxy is a thin disc which is filled with sufficient dust to restrict our view across it. The last naked eye visible supernova was Kepler’s Supernova in 1604, at a distance of about 6 kpc. (The parsec, a customary astronomical unit of distance, will be used here. 1 pc ~ 3.26 light years.) Despite the large distance (our Galaxy is only about 20 kpc in diameter), it was at its peak the brightest star in the sky. This is because, like most supernovae, it puts out energy equivalent to that which the Sun would produce over its entire lifetime, in a single burst. Over long, geological timescales, the Earth has had good odds of one of these going off quite close, with probable serious damage. Both photons (UV and X-ray) and cosmic rays are issues.

*Gamma ray bursts* have only been observed in other galaxies, not in our own. Their existence has only been known for a few decades. Amazingly, they were first detected when a satellite in near-Earth orbit designed to detect above-ground nuclear tests instead picked up an extragalactic signal. There is an active research program to understand them, but it is still (barely) possible to describe them in a single review: Gehrels et al. (2009). It is thought that their total energy is not a great deal larger than that of a supernova, but since it is channeled into (probably two) relatively narrow beams, significant damage in the form of radiation sufficient to trigger extinction-level ozone depletion could come from a distance of several kpc. In fact, they may be supernova-like events, with the energy unusually collimated into jets so that they appear more energetic if the observer lies within the direction of the jet. Rate and damage estimates suggest that they may well be a threat comparable to that of supernovae over long geological timescales. Gamma-ray photons are a definite issue; GRBs have been advocated as a primary source of high-energy cosmic rays, which would also be an issue if the GRB were close enough for photons to constitute a danger. There are three main types of GRBs but we will find that only two appear to be important for possible terrestrial effects.

*Pulsars,* as might be expected, pulse. They are rotating neutron stars, the remnants of past supernovae. The source object is very small, as small as 10 km in diameter, but with the mass of a star. They are remnants of supernovae, and are presumed to be solid “neutronium,” effectively one gigantic atomic nucleus held together by gravity. The observed pulses originate from a “searchlight beam” effect of electromagnetic radiation as they periodically sweep past. They were first detected in the radio, but emit in a wide range of frequencies. *Magnetars* are less well understood. They are strongly magnetized pulsars, given to occasional outbursts of gamma-rays. Enough of them are known to place some lower limits on the likely event rate at the Earth, as we do in section V.5. X-ray and soft gamma-ray photons are important.

*Active galactic nuclei* (called *Blazars* when we look “down the barrel” of the jet of emitted radiation) are rare, but might be sufficient to trigger extinction level events over a portion of a galaxy. All sorts of emission are possible, as they are a very heterogeneous population. They are typically powered by black holes of a million solar masses and up; our own galaxy is host to a largely dormant object near the lower end of this range. These large black holes episodically accrete matter in the form of gas or even whole stars. As this matter swirls down into the hole, it is strongly heated and may emit a great deal of radiation. This radiation is often concentrated in jets, because of magnetic fields and/or the fact that other pathways are blocked by the infalling gas, often arranged in a torus. There are many different kinds of AGN; the non-astronomical reader is most likely to have heard of quasars, among the most luminous AGN. External AGN may be a source ( Blümer et al. 2009) of the highest energy cosmic rays (which, however, are few in number). Our galaxy appears to be host near its center to a dormant phase of one of these objects. Since there is considerable dust and gas between the center of the galaxy and most of the disc stars, most of the radiation would be down-converted to a generalized infrared glow, probably harmless unless a jet were directed at the Earth. Jets are probably episodic, and the general rate of AGN activity has declined greatly over cosmological time. Haggard et al. (2010) conducted a study of the fraction of X-ray emitting AGN; about 1 in 600 are active, but most of those are much less powerful than quasars, and have insufficient emission (typically about as much as a supernova, but emitted about 8 kpc away near the dynamical center of our Galaxy. Quasars are extremely rare, so much that it is improbable that our Galaxy had a quasar phase any time since the Earth formed. We will not consider a possible outburst within our own galaxy, though there may possibly have been a significant outburst in the past.

*Galactic cosmic rays* are generally presumed to be fairly constant, except for fluctuations introduced by nearby supernovae (Erlykin and Wolfendale 2010). However, Medvedev and Melott (2007) suggested that there will be a long-timescale variation in cosmic-ray intensity modulated by the Sun's motion normal to the Galactic plane. This hypothesis was put forward in order to explain a strong 62 Myr periodicity in fossil biodiversity (Rohde and Muller 2005; Cornette 2007; Lieberman and Melott 2007, 2009; Melott 2008; Melott and Bambach 2011a,b)

*Diffuse backgrounds* exist for energetic photons. The X-ray background is thought to be primarily due to emission from million-degree hot gas in our galaxy. These backgrounds are orders of magnitude too low to have a significant terrestrial effect.

## V. **Types of Ionizing Radiation Events and the Terrestrial Fluence**

1. Strategy

Since direct effects of astrophysical ionizing radiation on the ground are poorly understood, we will only be able to comment qualitatively on them. The emphasis in radiation biophysics has been on direct effects of the sort of radiation produced by nuclear weapons, and that from cosmic rays incident on astronauts outside the atmosphere; both are different from shower secondaries which dominate on the ground. Some work estimating the effectiveness of the direct photon radiation on DNA damage (Martin et al. 2010) and photosynthesis productivity (Peñate et al. 2010) exists. Direct effects are more important for incident cosmic rays, which are deflected and diffused by the Galactic magnetic field. As discussed above and in the following sections, the indirect effects due to ozone depletion are a much more frequent and severe problem than the direct ones. It was shown in Ejzak et al. (2007) that the time duration and development of an irradiation episode is, from 0.1 s up to 3 yr, essentially irrelevant to the long-term maximum ozone depletion at the Earth. There is some dependence upon the energy of individual photons, with higher-energy photons producing larger ozone depletion for a given total radiation intensity. In some cases, such as supernovae, we will be able to give a rough estimate of the frequency of events of a given fluence at the Earth. (Fluence refers to the energy deposited per unit area.) In others, such as Solar flares, we will only be able to suggest upper limits and possible lines of future research.

A fluence of 100 kJ/m$^2$ at around 200 keV typical energy has been a standard for extinction-level ozone depletion in our past work (e.g. Thomas et al. 2005). Such an event results in about 30% globally averaged ozone depletion (Thomas et al. 2005; Melott and Thomas 2009). We expect this to be disastrous for the biosphere for several reasons. First, this value is a *global average*; local column density depletion can be as high 75%. For reference, the current globally averaged ozone depletion due to anthropogenic forcing is around 3-5%, with local maximum depletion of around 55% in the austral spring. Present day maximum depletion levels occur for a few months of the year, over very limited geographic extent (the southern polar region). In contrast, our high levels of depletion persist over much larger areas, at mid-latitudes, for many years.

We will normalize our discussions to the kind of depletion generated by such an event. We will scale to other fluences and energy spectra using the results shown in Ejzak et al. (2007). We know that up to this fluence, the amount of ozone depletion scales very approximately as the cube root of the fluence. Our “bottom line” will be a “mean lethal interval” or upper limit on that time—how often the Earth may be exposed to such a depletion event. The cube root fluence dependence means that expected terrestrial damage varies only weakly with the intensity of a possible irradiation source, or with its rate (which determines how close the source is likely to be, which again sets the intensity of irradiation). Thus, the large uncertainties we will see in the rate-intensity relations of astrophysical radiation sources do not, for the most important effect, translate themselves into equally large uncertainties in the severity-rate relation for their effects.

Our damage estimates apply only to the Phanerzoic Earth, which is shielded from UVB by its stratospheric ozone layer. About 90% of the UVB incident on the Earth is stopped by stratospheric ozone, and the portion that reaches the ground is dominated by wavelengths >295 nm, which have much less effect on DNA (e.g. Häder 1997). The amount of ozone and its depletion by ionizing radiation is relatively insensitive to variations of the size documented (e.g. Berner et al. 2003) in the oxygen fraction of the atmosphere during this time. Results for stellar flares and stellar descreening would be radically different for stars of type different from the Sun (Smith and Scalo 2009). Effects and time-averaged rates over geologic timescales of most other astrophysical ionizing radiation events should be similar for any planet in our Galaxy with a terrestrial atmosphere. Supernova rates vary with galactocentric radius (Lineweaver et al. 2004). There should be enhanced probability of nearby supernovae and possibly cosmic ray backgrounds when the Sun passes through Galactic spiral arms, but there is no evidence that this shows up in the timing of past glaciations (Overholt et al. 2009).

For each category of event, we have much more information on the photons that are emitted than we have on any cosmic rays that are emitted. This is because, for the most part, we receive the photons relatively unimpeded (at least so long as the observing device is above the Earth's atmosphere). However, cosmic rays below $10^{15}$ eV have their paths strongly deflected in the Galaxy, as mentioned earlier. This means that we do not observe them directly, and could for the most part only guess about the origin of the cosmic rays. For example, it has for some time been thought that most of the medium-energy cosmic rays in our Galaxy are produced in supernovae, but only recently has this been confirmed (indirectly) by observing gamma-ray photons produced inside supernova remnants (Abdo et al. 2010).

## 2. Heliosphere Descreening Events

Normally the Earth is protected from the full force of galactic cosmic rays by the combination of its own magnetic field and a magnetic field embedded in a plasma "wind" streaming outward from the Sun. Pavlov et al. (2005a); see also Frisch and Mueller (2010) have described a scenario in which (irregularly, but on average about every 30 My) the Sun passes through a molecular cloud, a region of higher density, in its movement through the Galaxy, lasting for perhaps 1 My. Such events push the Solar wind back into the inner Solar System, likely exposing the Earth to the cosmic rays in the interstellar medium. Terrestrial magnetic field reversals happen at irregular intervals, typically 200 ky, so a magnetic field reversal is likely during such a descreening event. They estimate that the combination may produce 40% global average ozone depletion which should be disastrous for the biosphere (Thomas et al. 2005; Melott and Thomas 2009). Note that the Antarctic ozone hole sometimes reaches such a level, but this is an order of magnitude above the global mean depletion. The sun angle over the Antarctic region is quite low, so obviously ozone depletion over lower latitudes can be much worse. Plots of depletion as a function of latitude are shown in, e.g. Thomas et al. (2005).

Pavlov et al. (2005b) also note that the pressure may also destabilize the Oort Cloud and induce comet showers, and that dust from the cloud may reduce the amount of solar radiation reaching the Earth, and cause glaciation. The ozone depletion effect will cause UVB to reach the ground, and such an event would have UVB effects similar to radiation events, but with duration ~1000 y. We have no obvious way to estimate the rate of such events, when descreening and magnetic field reversal are simultaneous. Magnetic field reversals are far from regular in their timing. It has been noted (Smith and Scalo 2011) that this mechanism will not be a significant threat for any biospheres near M stars, which are low-luminosity. The habitable zone is so close to the star that descreening is all but impossible. Frisch and Mueller (2010) have recently discussed the interstellar medium near the Sun and the probability of encounter of the Solar System with various density phases of the ISM, and its consequences for the heliosphere.

## 3. Supernovae

Supernovae are spectacular explosions associated with late stages in the evolution of stars. It is thought that about two or three per century occur in our Galaxy, but most of them cannot be seen due to heavy obscuration, since we and most supernovae lie in the disc of the Galaxy. Supernova 1604, or Kepler's Supernova, was the last naked-eye supernova in our galaxy at a distance of about 6 kpc. Supernova 1987a at a distance of about 51.4 kpc was visible to the naked eye in a nearby dwarf galaxy, and was the only naked eye visible supernova in the era of modern instrumentation.

One can get an average rate by taking the total number of events per year in the galaxy, getting a rate "per star," and multiplying by the mean density of stars in a region at the galactocentric radius of the Sun. The long-term time *average* rate of nearby events is then about

$$R_{\rm SN}(\leq D) = 2 \text{ events/Myr} \left(\frac{D}{100\text{pc}}\right)^3$$

(valid for distances D < 100 pc; Fields 2004; see also Capellaro and Turatto 2001). This rate will vary over time greatly based on proximity to spiral arms, and more specifically to massive star-forming regions. At about 100 pc, the rate changes to a $D^2$ dependence, as the disclike geometry of the Galaxy comes into play (less populated volume lies within a sphere of radius D).

The most recent reasonably complete computational efforts on terrestrial supernova effects are Gehrels et al. (2003) for atmospheric ionization from radiation, and Fields et al. (2008) for pressure effects upon the heliosphere from the expanding blast shell (which can result in descreening as described earlier). Interestingly, both studies identify a distance of order 8-10 pc as critical for extinction-level ozone depletion from direct radiative effects on the one hand, and compression of the heliosphere to the radius of the Earth's orbit on the other. With this compression comes (1) increased exposure of terrestrial life to cosmic rays in general as well as from the supernova itself and (2) substantial direct deposition of radioisotopes on the Earth from the supernova remnant. So, in addition to simple increased intensity with proximity, a number of new dangers come into play at about this distance. Simultaneously, the deposition of substantial amounts of radioisotopes becomes likely, making it possible to verify such an event.

Atmospheric and descreening effects both suggest 8-10 pc as the threshold for disastrous (extinction-level) consequences from radiation, so about 1-2 events per Gyr at this level can be expected. Improvements in the rate estimate are possible, and there are a number of ways this might be done. First, there are two main types of supernova, with numerous subdivisions. We do not need to delve into this, except to use some individual rate information for the two types combined with more detailed information about their ionizing photon and cosmic ray emission to improve these estimates. The Gehrels et al. (2003) computations were based on information from Supernova 1987a, which was of an unusual type, and assuming it went off close to us. The cosmic ray background for the simulations was constructed by taking the background we receive now, and scaling up its amplitude. The spectrum of cosmic rays would of course be quite different from this, probably containing a higher fraction of high-energy cosmic rays, which would change the terrestrial effects. The spectrum has been modeled by Ptuskin et al. (2010), building on earlier work. It is also the case that there will be fluctuations and anisotropy in cosmic ray background even in the absence of an extremely near event, due to a superposition of discrete sources and complicated, energy-dependent propagation (Ptuskin et al. 2006; Erlykin and Wolfendale 2006). Work on computing the atmospheric effects of different CR spectra has only begun (Atri et al. 2010). Fortunately, there are also now theoretical CR spectra which have been partially checked against data (e.g. Ptuskin et al. 2010 and references therein). This study suggests that Type IIb supernova remnants produce substantially more energetic cosmic rays than other types, so future work will need to take into account differential rates by supernova type. In addition, CR are probably trapped effectively in a supernova remnant, and their diffusion out of the remnant can take ~ a few thousand years (Fujita et al. 2010). This means that there is a strong transition in effects at the distance about 10 pc: closer than this, the Earth is inside the remnant (Fields et al. 2008), and exposed to the full effect of the trapped cosmic rays up to $10^{15}$ eV; outside it the effects should be less severe, and damped over the thousand-year timescale (Erlykin and Wolfendale 2010).

The secondaries that reach the ground from a nearby event are probably dominated by thermal (slow) neutrons and high-energy muons. To date there has no work published on their expected intensities. Fields et al. (2008) and references therein contain some information about the effects of direct deposition of radionuclides on the Earth; effects of these will be better understood as a result of Cold War era studies related to nuclear war scenarios. The long-term photon emission of the remnant (hot expanding shell of gas) in the UV to gamma-ray range was (in Gehrels et al. 2003) scaled from SN 1987a; observations taking place now will soon improve the picture greatly (Bufano et al. 2009). The X-ray emission of remnants was reviewed by Immler and Lewin (2003), but Immler maintains an up-to-date list at this website: http://lheawww.gsfc.nasa.gov/users/immler/supernovae_list.html . In the gamma-ray regime, Abdo et al. (2010) have reported a sufficiently precise result to characterize the gamma-ray emission of a remnant and indirectly characterize the CR background that drives it. An archive of similar results will go far to give us information on conditions inside a remnant—where the Earth will be if a supernova is close enough. Lastly, a variety of computations suggest a flash of energetic (UV and X-ray) radiation at the instant a supernova goes off. Usually, supernovae are only discovered a few days after this, but we have one instance of catching the flash as it happens (Soderberg et al. 2008). Such high-energy photons have to be observed above the atmosphere, from satellites. Presumably observations of this type will continue, and as the data set improves, this can be coupled to improved computational tools to make more accurate estimates of the effects on the Earth.

Ellis, Fields, and Schramm (1996) showed that short-lived radioactive species can provide the terrestrial signature of a nearby SN. Knie et al. (1999, 2004) have shown evidence of an excess of $^{60}Fe$ in ocean floor deposits, estimated to be 2.8 Myr old. Theoretical efforts (Benítez et al. 2002) suggest that the best candidate for this SN source, based on distance estimators, is in the Scorpius-Centaurus OB association. Repeated supernovae in this area over the last 10 Myr may be responsible for excavating the local hot gas region known as the Local Bubble in the interstellar medium (Frisch and Mueller 2010). Most recently, it has been suggested (Bishop and Egli 2010) that the $^{60}Fe$ isotopic signal could be found in magnetotatic bacteria fossils, and an experimental effort is underway. Supernovae at distances of about 30 pc, typical of this association, are not candidates for a major extinction event, but might have initiated a moderate one and had measurable effects on the Earth's climate. The grouping from which the precursor probably came has now moved to about 130 pc distance.

Table 1 summarizes specific *estimates* of the fractions of the total number of supernovae and the ionizing radiation energies of various kinds of supernovae. To get the number per century in our Galaxy, multiply the fractions in the Table by about 3. To get the radiation fluences we use later, we construct a weighted average of the sum of the burst energy and the total energy of the light curve phase. We emphasize that much of this information, particularly on UV, X-ray, and gamma emission early in the event will be improved by observations underway now. Much of the cosmic ray information is indirect, and again is improving with current gamma observations of supernova remnants. When we construct a weighted average over the various types, we find that the assumptions of Gehrels et al. (2003) are not far off, given the uncertainties, in spite of the first-ever direct observations of Soderberg et al. (2008). On the other hand, Gehrels et al. appear to underestimate the potentially severe effects of UVB, and to underestimate the cosmic ray energy with respect to modern rate estimates. We suggest that the "lethal distance" should be revised to ~10 pc, leading to a rate of 2 per Gyr or higher—a slight upward revision, with strong episodic rate increases when the Sun passes near starforming regions such as spiral arms. Note that this distance is larger than would be estimated from the photon effects alone, as the Gehrels et al. (2003) model has more total energy in cosmic rays than photons over the 20 year period of their computations. The effect of cosmic ray fluctuations is only beginning to be understood (Erlykin and Wolfendale 2010). Future astronomical observations in the X-ray and gamma-ray regime will help to refine this estimate.

Table 1: Approximate parameters of the main types of supernovae

| SN type | Probable fraction of SNe in Milky Way Galaxy (Ptuskin et al. 2010; see also Capellaro and Turatto 2001) | Burst energy (time in seconds) Results are preliminary. | Cosmic ray spectrum cutoff (theoretical --Ptuskin et al. 2010) | X-ray power from light curve phase (weeks/months) (NASA GSFC list of X-ray supernovae) |
|---|---|---|---|---|
| Ia | 0.32 | <$10^{39}$ J (2s) in X-ray and gamma-ray (Hoflich and Schaefer 2010) | $10^{16}$ eV | ~$10^{31}$ J $s^{-1}$ at $10^7$ s |
| Ib/c | 0.22 | 2 ×$10^{39}$ J (~129s) in X-ray < 100 keV (Soderberg et al. 2008) | $10^{16}$ eV | 2 X $10^{33}$ J $s^{-1}$ (~$10^4$ s) Evolves as $t^{-0.7}$ (Soderberg et al. 2008) |
| IIbc | 0.02-0.04 | Unknown—likely large fluence | $10^{18}$ eV | $10^{32}$--$10^{34}$ J $s^{-1}$ Highly variable |
| IIP | 0.44 | ~$10^{37}$ J (few hours) in UV (Schawinski et al. 2008) | $10^{15}$ eV | $10^{32}$--$10^{33}$ J $s^{-1}$ at $10^4$ to $10^7$ s |

We wish to determine the frequency $f_\delta$ of events at a given ozone depletion level δ, with ozone depletion serving as a proxy for overall damage. The fluence received scales with distance D from the supernova as $D^{-2}$. As noted above, past work indicates that ozone depletion scales approximate as the cube root of the fluence. Therefore, ozone depletion will scale as $D^{-2/3}$. For relatively nearby supernovae at distances D ≤ 100 pc, smaller than the thickness of the Galactic disc, the probable number within a given volume varies approximately as $D^3$ (as shown in the rate equation above). With this scaling, then, the frequency $f_\delta$ of events at a given ozone depletion level $\delta$ will vary as $f_\delta \sim \delta^{-9/2}$. For distances D > 100 pc, the Galaxy becomes more like a thin disk and therefore the probable number scales as $D^2$. For this case, then, $f_\delta$ will vary as $f_\delta \sim \delta^{-6/2}$. In Figure 1, we plot the rate of supernovae as a function of fluence received. For D ≤ 100 pc, this rate varies with fluence, F, as $F^{-3/2}$, while for D > 100 pc, the rate varies as $F^{-1}$. Further discussion of this figure can be found in section V. 8. below.

Taking 10 pc as the distance at which we would expect an extinction level event, the expected rate would be about 2 $Gyr^{-1}$. This rate will rise rapidly for less severe events. It will be about 1 $Myr^{-1}$ for events which introduce about 10% ozone depletion, which is greater than current anthropogenic effects on a global scale and would therefore be a noticeable stress on the biosphere. The amount of damage will reach extinction level for the rare very nearby events, roughly ≤ 10 pc (see Figure 1), for which the Earth can be inside the supernova remnant, subjected to much higher cosmic rays and direct irradiation.

Even for events not near enough for this catastrophe, past events proximate to our Solar System may in principle be detected by radionuclides. Given that there is ice on the Earth up to a maximum age of about 8 Myr, additional evidence may be collected in core samples, although the oldest continuous core samples are a few hundred thousand years. Such an event should (for nearby supernovae, not historical ones) deposit nitrates from the ionization (Thomas et al. 2007), along with radionuclides but not an ammonia spike, which can be taken as evidence of a bolide impact (Melott et al. 2010b). More distant supernovae may produce detectable $^{14}C$ in the atmosphere, and estimates of supernova gamma-ray emission which are reasonable in view of later work have emerged from work on ice core data (Kai-mei and You-neng 1988). This suggests additional work may be possible to constrain high-energy events near the Earth by ice core data and/or $^{14}C$ data.

4. Gamma-ray bursts

(a) GRB scaling

The large energy of gamma-ray bursts (GRBs) can make them a threat at considerable distance, as noted by Thorsett (1995), Dar et al. (1998), and Scalo and Wheeler (2002). What was thought to be an implausibly large energy is now understood as a beaming effect: we only “see” them when the energy is directed at us. The implication is that there are many more which we don't see, but since we are concerned with irradiation effects, we can scale directly from the observations, with some caveats. Therefore, when we quote rates, it will be without a beaming correction. The reader should understand that there are approximately 50-100 times (determined by indirect evidence) more than we observe and quote here as a rate, but that these are probably not important for terrestrial effects, except possibly occasional modest increases in the high-energy cosmic ray background.

Gehrels et al. (2009) recently reviewed the state of observational knowledge. It is important to note that GRBs are rare, extremely energetic events. Although earlier estimates suggest that the probability of a strong irradiation may be comparable to that from a supernova, the GRB results contain an additional extrapolation: all of them have been observed in the Universe at large, in which there are a few per day that are detected. So the scaling to a local estimated rate contains an additional uncertainty, which is the extent to which rates in an external galaxy will resemble rates in our own Galaxy. Kusenko (2010) has shown that the most straightforward way to account for the composition-energy relation of ultra-high energy cosmic rays is to assume a component from past GRBs in our own Galaxy.

Before exploring the question of rate extrapolation, it is worth examining its consequences. Based on previous work, we can make reasonably reliable estimates of how effects on the Earth vary with the fluence received. Thomas et al. (2005) found that the production of nitrogen oxides was nearly linear with fluence, but due to feedback mechanisms in the reaction rates (see Ejzak et al. 2007) the percentage ozone depletion scales much less than linearly—approximately as the cube root of fluence. This statement will only be valid up to fluences of a few hundred kJ $m^{-2}$, since above that level there is enough heating to invalidate the atmospheric code used to do the modeling. Fortunately, such events are very rare. The consequence of this sublinear scaling is that uncertainties in GRB rates do not translate into correspondingly large uncertainties in probable terrestrial damage levels.

When one examines GRB rates, it is clear that the distance D of the nearest probable event over any reasonable timescale is likely to be large compared with the thickness of the Galactic disc, so that the number of possible precursor stars scales as $D^2$. Since radiation intensity scales as $D^{-2}$, for a given GRB rate per star of *n*, the frequency at a given fluence scales as *n*; the frequency at a given ozone depletion damage level $f_\delta$ scales as $\delta^{-3}$; and most importantly given the rate uncertainty, $f_\delta$ scales as $n^{1/3}$. In Figure 1, we plot the rate of two types of GRB as a function of fluence received. For both types, the rate varies with fluence, F, as $F^{-1}$, but with a difference in overall scaling, as described below.

There will be some bias factor *b* ≤ 1 on the GRB rate given the type of galaxy in which we reside. We think that *b* ≤ 1 because there is some evidence that the host galaxies of the most powerful GRBs are somewhat different than our own: possibly smaller, lower-metallicity (translation: a lower fraction of elements other than hydrogen and helium), and with higher specific star formation rates (Stanek et al. 2006). We will revisit this factor with respect to each GRB type.

In addition to the burst emission, GRBs have an afterglow in the X-ray. This afterglow contains, in all except a very few exceptional cases, a much lower (typically 0.01) amount of energy than the initial burst (Oates et al. 2009). As such, since it will be irradiating an atmosphere with reduced ozone, it might contribute a factor of order unity to the short-term damage from the UV of the parent star (in our case the Sun).

There are two main categories of GRBs: long-soft GRB (LSGRB) with a lower spectral energy peak, longer duration, and higher total energy, and short-hard GRB (SHGRB) with opposite mean properties. There is evidence for a third intermediate class (e.g. Horváth et al. 2010). However, for our purposes this is a low-energy, spectrally soft extension of the LSGRB class, and ignorable with respect to terrestrial effects.

(b) Short-hard gamma-ray bursts

Short-hard GRBs (SHGRB) which have burst durations usually 2s or less and emit more "hard" photons, that is higher energy ones, probably result from the merger of black holes and/or neutron stars. SHGRB do not have a strong bias that would affect the rate in our galaxy, but they display some preference for older stellar populations. Based on Nakar (2007), we could set *b ~ 0.5*, but more recent results would suggest *b =1* (no bias; Cui et al. 2010). We assume typical peaks in the spectral emission around 800 keV, with energy of about $10^{43}$ J, and a rate of about 40 $Gpc^{-3}$ $yr^{-1}$. It is important to note that there is wide variation, and events with peak energies of a few MeV have been observed (Ackerman et al. 2010) Approximating the number density of galaxies as about $5 \times 10^{6}$ $Gpc^{-3}$, we can estimate a (biased) SHGRB rate of about $8b \times 10^{-6}$ per year in a galaxy like ours. A collimation factor was never included, so if there is no preferred jet axis, as seems reasonable, this is the number directed at us from somewhere within our galaxy—one every 125,000 $b^{-1}$ years. From our previous work, we know that "harder" spectra, i.e. those with more high-energy photons, cause a greater effective ozone depletion, because they penetrate deeper into the stratosphere, where the primary ozone layer lies. Scaling from Thomas et al. (2005) in terms of fluence and compensating for spectral hardness based on Ejzak et al. (2007), we put the "lethal" distance (again, arbitrarily set at ~30% global average ozone depletion) at about 200 pc. This is considerably closer than the corresponding distance for Long-soft (LSGRB) bursts (about 2 kpc), which were previously emphasized. Only about $4 \times 10^{-4}$ of the area of the galactic disc lies within this distance, but the SHGRB rate is higher, so the rate of "lethal" level events from SHGRBs is something like one per $300b^{-1}$ million years, where b is of order unity. It must be emphasized that this is an order-of-magnitude estimate, based on uncertain numbers scaled from observations with incompletely understood instrumental biases. Still, it suggests that short bursts are a new potential source of stress to the biosphere not previously considered.

(c) Long-soft gamma-ray bursts

Guetta, Waxman, and Piran (2005) suggest a rate of ~0.5 $Gpc^{-3}$ $yr^{-1}$ for LSGRB, but based on more recent (SWIFT, see Gehrels et al. 2009) data and based directly on data with a minimum of theoretical models, Wanderman and Piran (2010) estimate the global rate (with no beaming factor) as 1.3 $Gpc^{-3}$ $yr^{-1}$. Such events have typical spectral energy peaks around 200 keV, and a typical total energy around $5 \times 10^{45}$ J. Note that the greater effectiveness of harder photon spectra in ionizing the atmosphere (Ejzak et al. 2007), and the correlation of spectral hardness with total luminosity (Gehrels et al. 2009) may make this a slight underestimate, but a more detailed treatment must await a more complete luminosity function, including exploration of the high-luminosity tail. This kind of event, which apparently happens in our Galaxy every few million years (pointed at us), was modeled in detail in Thomas et al. (2005); the "lethal" distance for a photon event is about 2 kpc (roughly 1/10 of the diameter of the Galaxy). This gives a mean interval of 110 $b^{-1}$ Myr for such events. The big uncertainty in this estimate is b. As emphasized in Stanek et al. (2006), there seems to be a bias in such burst types toward low-metallicity environments. More recent work has suggested a host of other environmental preferences, including active star formation. This type seems to result from core collapse to a black hole of a very high-mass star with high angular momentum, and it has been argued that the low metallicity is necessary for formation of an LSGRB (e.g. Woosley and Zhang 2007). There is a large literature on this question, and it is inappropriate to tackle its complexities here. Levesque et al. (2010a) have examined it, and conclude that a host galaxy metallicity bias in the rate of LSGRBs exists. They emphasize that it is not a cutoff, and give examples of high-metallicity environments that have hosted LSGRBs. Levesque et al. (2010b) find that there is no similar bias in the energy of LSGRBs. There are indications that in some cases, the optical photons are obscured by dust, which makes the burst site less optically visible. The evidence listed there as well as work cited suggests values around $b \sim 0.1$ are reasonable for our Galaxy, and that we can assume that the energy of any LSGRB bursts would be representative of the Universe as a whole. For a summary of these results see Levesque (2011).

There is a probable additional population of low-luminosity GRBs, which have approximate parameters 325 $Gpc^{-3}$ $yr^{-1}$, and energies of about $10^{40}$ J (Liang et al. 2007). These are probably not a low-energy tail of the normal LSGRB population. The total energy of these events is not too different from that of a supernova, but their rate is apparently about 3 orders of magnitude lower than the supernova rate. Consequently we ignore them for purposes of our total rate estimates. Note also that their existence was partly postulated based on numbers in excess of the low-luminosity extrapolation of the LSGRB luminosity function, so we find that our overall estimates will not be sensitive to a cutoff at this end of the function.

Our assignments of values for b are very tentative, but using 1 for SHGRBs and 0.1 for LSGRBs, we estimate the total rate of “lethal” events irradiating the Earth from gamma-ray bursts of both kinds to be approximately 4 $Gyr^{-1}$. In addition to the tentative assignment of rates and b values, we emphasize that lethality means serious damage to biosphere, possibly extinction-level. It is obviously impossible to collect experimental data on entire ecosystems irradiated with approximately doubled UVB fluxes, but all the data on individual organisms suggests extremely serious effects (see discussion and citations in part IV above). Further data from the variety of high-energy orbiting observatories will clarify the threat. Hopefully the information presented here will enable some scaling to new data as it emerges.

(d) Cosmic rays—special issues for GRBs

The cosmic ray content of GRB jets is unknown, but there are strong arguments to suggest that it is very large. It may even be competitive with the energy in photons. It has been argued that GRBs may in fact be the source of nearly all cosmic rays with energies above about $10^{15}$ eV (Dermer 2002). Again due to magnetic fields the propagation of such cosmic rays within the galaxy is complicated, but it is inescapable that if there is a substantial proton loading in the jet, a strong burst of very high-energy cosmic rays $\geq 10^{15}$ eV will punch through the magnetic field and strike immediately after the photons (Dermer and Holmes 2005). There will be a continuously changing residual excess detectable for a long time—possibly several thousand years. Such an excess has been proposed to explain anisotropy and energy spectral features in recent high-energy cosmic ray data (Calvez and Kusenko 2010; Calvez et al. 2010). This baryon loading would significantly exacerbate the stressful effects of a GRB on the biosphere, increasing the atmospheric ionization and adding a substantial load of high-energy muons and thermal neutrons to irradiation of organisms on the ground and the surface of sea, with substantial muon flux reaching 1 km below the ocean surface. We are only beginning to explore the terrestrial effects of very high energy cosmic rays (Atri et al. 2010). Detection of neutrinos coincident with GRBs is expected give information on cosmic ray loading in the near future (Kappes et al. 2010).

## 5. Pulsars and Magnetars

Pulsars are rapidly rotating neutron stars which emit radiation across the electromagnetic spectrum. Typical power is $10^{27}$ to $10^{29}$ watts. This is of course greater than the power of the Sun ($4 \times 10^{26}$ watts), but still these are not a serious threat. The self-healing capabilities of the terrestrial atmosphere can establish a reasonable equilibrium with about 1 mW $m^{-2}$ of X-rays or gamma-rays (Ejzak et al. 2007), which is close to the Solar X-ray emission at Solar maximum. In order to have even a moderate effect, the most powerful known pulsar would have to be within 0.1 pc, far smaller than the mean distance between stars.

Magnetars have extremely powerful magnetic fields and are prone to powerful outbursts; such objects are referred to as soft-gamma repeaters. One such recent event (Palmer et al. 2005) was estimated to have an energy of $10^{39}$ J, though more recent estimates (Bibby et al. 2008) of the distance suggest that this should be lowered to $3 \times 10^{38}$ J, with a crudely estimated rate of such events of one per 30 years in our galaxy. Amazingly, for the first time the signature of an extrasolar ionizing event was detected in real time. Mandea and Balasis (2006) reported an increase in high-altitude atmospheric ionization of six orders of magnitude. This in turn produced noticeable changes in the geomagnetic field and modified radio transmission. The “lethal distance” estimate for this energy is ≤ 0.5 pc; given our crude rate estimate this should occur less than once in the age of the Earth. So, these are ignorable, except for having given us a fine example of real, measurable terrestrial effects from an event halfway across the Galaxy.

## 6. Solar Flares and Solar Proton Events

Events on the Sun have great potential to affect the Earth, through both photon and proton channels. Solar flares emit electromagnetic radiation from the radio through visible light, UV, and X-ray. There are Solar proton events (SPE)—some of which involve a small mass of protons with high energy, and coronal mass ejections (CME) which involve a large outpouring of lower energy protons toward the Earth. One of the main differences with the other events we have been considering is that we have more data, due to the proximity of the Sun and the frequency of moderate intensity events. Fast CMEs can impact the geomagnetic field and cause it to pulse, then travel along field lines into the atmosphere, largely in polar regions. There is considerable concern because such events have already (e.g. 1989) caused widespread power outages. The increasing dependence of our civilization on electronic devices and possible economic damage in the vicinity of $100 billion has caused considerable concern, recently reviewed in a National Research Council (2008) report.

The largest historical SPE (called the Carrington Event) took place in 1859, with spectacular results, reviewed in the volume edited by Clauer and Siscoe (2006). The emerging telegraph industry was affected by induction in the wires—with a few cases of fires being set in telegraph offices by the resulting current surge. The *aurora borealis* was seen in Jamaica and Hawaii. This event was several times stronger than any subsequent event, and probably the strongest in the last five hundred years. The atmospheric ionization of an SPE produces oxides of nitrogen that are recorded in ice cores. It has been possible to model this process and produce reasonably accurate agreement with the ice core data. Thomas et al. (2007) and Rodger et al. (2008) have modeled the atmospheric effects of the 1859 Carrington Event. Other events have been studied through both satellite observations and computational modeling; for a review see Jackman and McPeters (2004). The terrestrial impact of X-ray emission in such events is probably negligible in comparison to the proton fluence (Thomas et al. 2007).

There is, therefore, considerable interest in the long-term rates of powerful SPEs. Fortunately, it has been possible to get information on impulsive events from nitrate deposits in ice core data, and limits on integrated fluence over long timescales in lunar rocks (e.g. Smart et al. 2006). $^{14}C$ is produced in the atmosphere and lunar soil as well. This and other radioisotopes can be used to constrain the magnitude of impulsive events (using tree rings and ice core data) and integrated total fluence (using lunar soil data). All of this is summarized in a very nice review (Usoskin 2008).

The results, which have not changed recently, are summarized in Figure 6 of Smart et al. (2006). They are presented there in terms of events per year with a fluence of protons >10 MeV. We reproduce and extend these results in our Figure 1. Events with a fluence F of $10^7$ $cm^{-2}$ of such protons occur about 10 times per year. The rate continues as a power law $F^{-0.4}$ out to about $10^{10}$ $cm^{-2}$ , where it breaks. The Carrington Event had an estimated fluence of about 3 × $10^{10}$ $cm^{-2}$, and according to ice core data has not been matched in the last approximately 500 years.  This fluence was converted to energy, and the resulting computations match well to ice core data, when the nitrate deposition is taken as integrated over a few months (Thomas et al. 2007; Melott et al. 2010b).

Ice cores and $^{14}C$ (both terrestrial and lunar) give consistent results back some few times 10,000 years suggesting that the power breaks to $F^{-0.9}$ at about $10^{10}$ $cm^{-2}$. This is also consistent with limits from lunar data on integrated fluence from concentrations of radioisotopes of Ca, Kr, Al, and Be at the million-year level.  It is important to note that these upper limits on long timescales are very weak. These upper limits correspond to fluences so large that they might end all life on Earth. Also, they are really upper limits for many kinds of events, since many things other than Solar events might cause such deposition.

There is one "suspicious coincidence" suggesting a possible event in the not too distant past. The climatic transition around 10,000 years ago from "ice age" conditions to modern, more temperate ones was accompanied by a 500-1000 year return to ice age conditions, called the Younger Dryas. (a) This was accompanied by a moderate extinction event and a strong upswing in ice core nitrate levels, which could be radiation induced. This nitrate event has been blamed on widespread fires due to climate drying, as well as a possible comet impact. Melott et al. (2010b) showed that the nitrate peak corresponded to the injection of about 100 "Carrington events" worth of ionization into the atmosphere; the Carrington event produced about five years' worth of nitrate at recent "normal" deposition rates (Dreschhoff 2003).  (b) Objects are dated using $^{14}C$, which decays but is continually produced in the atmosphere by cosmic rays. There is also a $^{14}C$ excursion around this time. There is a well-known problem in dating in that ages determined from $^{14}C$ are consistently younger than ages determined other ways (for a  review see Fairbanks et al. 2005), assuming that the $^{14}C$ production rate matches the current rate. This can be explained by many things including changes in ocean circulation, and changes in the geomagnetic field which let in more cosmic rays than we see recently. It is nearly always tacitly assumed that the "cosmic" background is constant.  Of course it need not be, and the slope of the age versus $^{14}C$ age curve implies a long term average production rate over the last 50,000 years about 10% higher than we currently see, with fluctuations corresponding to about 1000 years of $^{14}C$ production. It is interesting, although not compelling, that a "100 Carrington" ionizing event could reasonably change the climate, fit the nitrate and the $^{14}C$ excursions, and fit just below the upper limit curve we have described. More research is needed on chemical and geoisotope excursions in this recent era, when we have access to ice cores. The oldest ice known on Earth is about 8 million years old (Bidle et al. 2007) though systematic isotope and/or chemical analysis is so far restricted to less than a million years (Barbante et al. 2009).

We have contented ourselves with a verbal description of Figure 6 of Smart et al. (2006), because for our own Figure 1 we convert the fluences to SI energy units and compare them with likely fluence-frequency distributions of other kinds of events (supernovae, GRBs, etc) that have been suggested here. In doing so, it must be clear that many of the photon events would produce nitrate (it only takes about 10 eV to break the nitrogen molecule triple bond) but nuclear reactions typically demand MeV energies, about 10 MeV for $^{14}C$. Consequently the isotope constraints from lunar soil will not apply to many of the photon events which would be devastating to the biosphere.

## 7. Backgrounds and possible periodic cosmic ray enhancements

This review will not consider the long-term effects of backgrounds, as the focus is on unusual enhancements of the rate. Juckett and Rosenberg (1997) and Juckett (2007, 2009) have discussed evidence correlating modest enhancements of the cosmic ray background with cancer.

Medvedev and Melott (2007) have described a scenario in which enhanced cosmic rays associated with the motion of the Sun in the Galaxy may explain a 62 Myr periodicity in fossil biodiversity. While the periodicity is found in multiple data sets and using a variety of analysis methods, (Melott and Bambach 2011a), the proposed mechanism depends upon the unknown cosmic ray generation at the Galactic bow shock and the amount of shielding provided by the turbulent Galactic magnetic field. The viability of the mechanism depends upon these unknown parameters (Melott et al. 2010a). It would be premature to include this here without further research.

In addition, Melott and Bambach (2010) have confirmed a 27 Myr periodicity in extinction. Once again, such long timescale periodicity with high regularity suggests an astronomical connection, but at the moment there seems to be no viable astronomical mechanism for this. As for the mechanism of Pavlov et al. (2005a) described in section V. 2., we can only estimate a fairly wide range of possible severity.

### 8. Comparison of rates and intensities

Our approach is borrowed from that used in, e.g. Smart et al. (2006) to display constraints on Solar proton events. In Figure 1 we show a plot of rates (per year) versus fluences for impulsive events. The two vertical lines show approximate fluence thresholds for (1) noticeable ozone depletion similar to the current anthropogenic damage, with measurable biological impact and (2) an extinction level depletion, with doubling of the global average UVB fluence on the surface which should seriously impact the marine food chain.

It is important to note that in presenting this plot, we have made an approximation and simplification in the presentation of proton fluences. This is done for three reasons: (1) For most objects other than the Sun, we don't know what the proton fluence would be. We can only approximate its total energy or its spectral distribution, and only have approximate understanding of its probable time development. (2) Many of the effects of cosmic ray protons and other nuclei have not yet been elaborated. The "radiation on the ground" enhancement requires extensive air shower simulations which are just now getting underway (e.g. Atri et al. 2010; Melott et al. 2010a). (3) Most energy from cosmic rays as well as ionizing photons is deposited in the atmosphere, and does not reach the ground. The dominant effect for low to moderate events will be ozone depletion, just as in the proton events, and work to date (e.g. Gehrels et al. 2003; Thomas et al. 2005) shows that using the "energy deposited" approximation (Ejzak et al. 2007) works well in describing the ozone depletion. So we will use this approach, and describe how various estimates would change if cosmic rays were included. *That change will be a minimal estimate that does not capture additional short-term and long-term effects on the biosphere.*

For Solar proton events, we have converted the proton fluences of Smart et al. to energy fluences using 30 MeV per proton, which is a reasonable procedure at the level of approximation we are working with. The shallow solid line on the upper left corresponds to our direct knowledge of Solar proton events. That curve inflects in the vicinity of 200 J $m^{-2}$, with a few points inferred from nitrate levels in ice core data, and a dotted line connecting weak upper limits at very high fluences from radionuclides on the Moon.

Clearly, the low-fluence events are dominated by the Sun. Above the fluence of the Carrington event, the largest known SPE, there is no data beyond the very weak upper limits. Potentially, the Sun could deliver a serious blow to the biosphere, and as emphasized by the NRC report, to our technology. Schaefer et al. (2000) identified superflares many orders of magnitude larger than the Carrington Event on a variety of Sunlike stars. However, Rubenstein and Schaefer (2000) described a mechanism wherein close-orbiting superjovian planets become tangled in the magnetic field of the star, and a reconnection event releases large amounts of energy (for a narrative description, see Rubenstein 2001). If they have correctly identified the mechanism, this is not a process operative in our Solar System. It might well make life impossible in systems with close-in superjovians, however. More generally, due to the hazard level and possible frequency, it would be worthwhile to try to get more information on the shape of the cutoff. The × mark on Figure 1 suggests a region of parameter space where it might be possible to get an additional limit on Solar activity by combining ice core data with atmospheric irradiation simulations (e.g. Thomas et al. 2007; Atri et al. 2010). Certainly, upper limits much stronger than those from lunar data are possible.

The fact that life exists on Earth indicates that events too far to the right of the extinction line are not too frequent. On the other hand, incomplete extinction events are common and have happened as recently as 12,900 years ago. Extremely severe events called mass extinctions happen less frequently; Bambach (2006) has identified 19 over the last 550 My. Radiation events may play a role in these. Melott and Thomas (2009) compared some of the characteristics of the end -Ordovician extinction event with those expected for astrophysically induced ozone damage, finding consistency. The challenge is to find one or more “smoking guns,” because most such events are “clean,” leaving few traces. (For a possible exception, see Athanassiadou and Fields 2011.) While the Solar curve is based on direct detection or upper limits from data, the other curves are constructed from the estimates made earlier.

We have estimated curves for long-soft and short-hard GRBs (long dash and dot-short dash lines), and for supernovae (long dash-short dash). Pulsars and magnetars are too low in fluence to pose a probable threat, though as noted one event has had a measurable, though small, effect on the ionosphere. These curves are estimates of probable rates based on astronomical data.

Supernovae and both varieties of gamma-ray burst occupy the lower portion of the plot. The supernova curve (shortdash-longdash) has two inflection points. There is one inflection (at about $10^5$ J $m^{-2}$) for events that are so close that the Earth is inside the active supernova remnant, with the solar wind compressed inside the Earth's orbit, and gets an extra blast of high-energy cosmic rays. The other inflection (at about $10^3$ J $m^{-2}$) is for events that are sufficiently distant that the distribution of contributing events takes on the disclike structure of the Galaxy. The rates are low, but not negligible over the geological timescale. Extinction level events are probable every few hundred million years. Because we assume the rate of long-soft gamma ray bursts in our Galaxy is one-tenth of the cosmological mean, the LSGRB line is the lowest. Supernovae and SHGRBs appear to dominate the rate of extinction-level events.

GRB events with a fluence below about 1000 J $m^{-2}$ (for LSGRBs) or 10 J $m^{-2}$ (for SHGRBs) have rates which are hard to estimate and probably quite low. These limits correspond to irradiation from the maximum distance within the Galaxy. A burst from Andromeda would be inconsequential for the biota, so such low levels could be reached (for example) by a narrow miss in the fringes of the beam, or a burst originating in one of the dwarf galaxies which orbit the Milky Way, or a burst of the rare "third type" (Horváth et al. 2010). Consequently, although there is some unknown probability of events at low fluence, we cut off the lines at fluences corresponding to origination of typical GRBs within the Galactic disk.

It should be noted that the LSGRB curve assumes a bias of 0.1, and should be moved upward one order of magnitude in rate if there is no such bias. Either kind of GRB curve could be moved to the right a factor of two if beams contain cosmic ray energy comparable to the photon energy, and additional effects would exist to damage the biota. The supernova curve contains a large inflection for very close encounters which is designed to capture a minimal estimate of the disastrous effect of being inside a supernova remnant. All of the cosmic ray effects, if present, could be greater than we have indicated.

There are significant uncertainties in these three rate estimates. In the future, it should be possible to refine them as new information emerges. Ground-based high energy cosmic ray detectors and orbiting observatories are gathering new information on the rate and nature of high energy events. This has been going on only a few decades, and at least one unanticipated new threat has emerged, in the form of gamma-ray bursts, at a level comparable to that of supernovae. Better information will enable us to better understand the development of our biosphere, possible effects on life in other kinds of planetary environments, and the near-term danger to our own technological civilization.

## VI Future Research

A great deal of improvement in the estimates of the global rates and the rates likely in our Galaxy for gamma-ray bursts will result from observations now taking place. A better understanding of GRB progenitors will enable a more accurate understanding of their probability within our galaxy. Multiple NASA and ESA observatories above the Earth's atmosphere provide information on the interior of supernova remnants, which provides information on their cosmic ray density and energy spectrum. Further research on magnetotatic bacteria provides an opportunity to confirm the presence of $^{60}Fe$ from a relatively nearby supernova 2.8 Myr ago. There have been suggestions that isotopic anomalies in the recent geologic past may arise from supernovae. Estimates (A. Overholt, personal communication) suggest that this is a reasonable possibility. Since these anomalies may arise from a variety of ionizing radiation sources, they can be used to set upper limits on solar or supernova events.

We are working on producing tables which will enable the computation of the amount of high-energy muons and very low-energy neutrons, which are likely to be the primary direct radiation hazards from enhanced cosmic ray flux. The tables can be combined with the energy spectrum of a particular source to compute a profile of the radiation flux on the ground which would result.

We are working with P. Neale (http://www.serc.si.edu/labs/photobiology/index.aspx ) to profile the UVB sensitivity of *Synechococcus* and *Prochlorococcus*, two picophytoplankton species that are abundant in the oceans and thereby make up a large fraction of the global primary productivity. These results will be combined with new and existing simulations of atmospheric chemistry and complete radiative transfer calculations of surface irradiance to improve understanding of impacts on the biosphere. In general, past UVB work has only explored the consequences of the small to moderate UVB increases expected from anthropogenic activity. It will be useful to know how this extrapolates into the more extreme region possible from astrophysical events.

Exploration of the characteristics of past mass extinctions will help to assess when radiation events have had an important role. The most important and most difficult feature of this is finding “smoking guns.” These events do not leave obvious impact craters! They are generally quite “clean,” with energy absorbed in the atmosphere, and leave few geoisotopic or geochemical traces. This is the extreme challenge of trying to piece together a past which must have happened in some fashion, but whose details are not at all clear.

### Acknowledgments

We thank C. Dermer, B. Lieberman, and G. Rudnick for helpful comments. We thank our collaborators and other colleagues, too numerous to list here, from whom we have learned bits and pieces of the many kinds of science that join here. Research support was provided by the NASA Program Astrobiology: Exobiology and Evolutionary Biology under grant number NNX09AM85G.

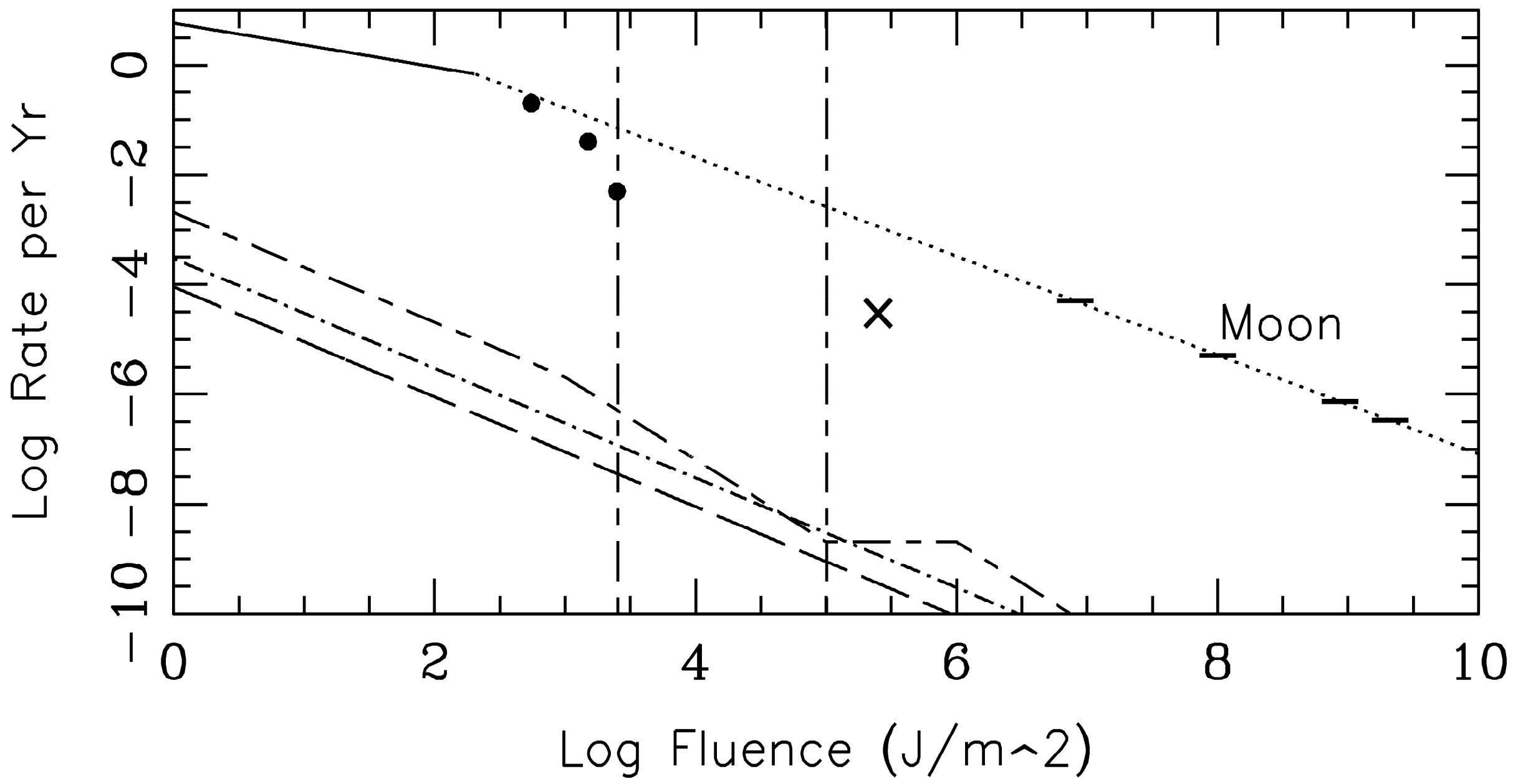

**Figure 1:** Rate (per year) estimates and limits for various types of astrophysical ionizing radiation events that may impact the Earth, as a function of their fluence at the Earth in Joules per square meter. The solid line at the upper left is based on actual measurements of Solar Proton Events in recent decades. The solid circles are based on SPEs over the last few hundred years with parameters indirectly inferred from ice core data, including the Carrington (1859) event. The horizontal lines labeled "Moon" are upper limits on cumulative exposure rates inferred from radionuclides in Lunar soil, and the dotted line connects this with the Solar data. It should, however, be considered an upper limit on all kinds of radiogenic events, including cosmic rays and photons above 10 MeV. The × indicates a conjecture on the sort of limit that should be possible from a complete survey of terrestrial isotope and ice core chemical data. The lower three lines are long-soft gamma-ray bursts (long dashes), short-hard gamma-ray bursts (dot-shortdash), and supernovae (longdash-shortdash). The LSGRB rate estimate assumes that the rate in our Galaxy is depressed by heavy element content; if this is incorrect it should be up one order of magnitude, becoming an equal contributor for extinction level events. Both gamma-ray burst curves assume that there is a negligible contribution from cosmic rays. If this contribution is instead substantial, they may be moved to the right by a factor 2 or so. Both gamma-ray burst curves are cut off at a minimum fluence expected from a normal burst originating within our Galaxy. The supernova curve shows an inflection at a fluence of about $10^5$ J $m^{-2}$. This exists because of a critical distance, inside which the Earth is exposed to the full cosmic ray load of the supernova remnant, without protection of the Solar wind. A mild inflection at $10^3$ J $m^{-2}$ is due to the shape of the stellar distribution in the Galaxy. The two vertical lines show approximate fluence thresholds for (1) noticeable ozone depletion similar to the current anthropogenic damage, with measurable biological impact and (2) an extinction level depletion, with doubling of the global average UVB fluence on the surface which should seriously impact the marine food chain.